%% file: main.tex
\documentclass{article}
\usepackage{spconf,amsmath,graphicx}
\usepackage{subcaption, multirow, multicol, booktabs,amsfonts, kotex}  
\usepackage[table,xcdraw,dvipsnames]{xcolor}



\input{command/mycommand}
\usepackage{bbm}
\usepackage{graphicx}
\usepackage{amsmath}
\usepackage{amssymb}
\usepackage{booktabs}
\usepackage{enumitem}
\usepackage{rotating}
\usepackage{colortbl}
\usepackage{makecell}
\usepackage{tabu}
\usepackage{subcaption}
\usepackage{multicol}
\usepackage{multirow}
\usepackage{csquotes}
\usepackage{wrapfig}
\usepackage{url}

\title{Learning Video Temporal Dynamics with Cross-Modal Attention\\for Robust Audio-Visual Speech Recognition}
\name{Sungnyun Kim\thanks{*\,Equal contribution.}$^{1*}$, Kangwook Jang$^{2*}$, Sangmin Bae$^1$, Hoirin Kim$^2$, Se-Young Yun$^1$}
\address{
    $^1$Kim Jaechul Graduate School of AI, KAIST \\
    $^2$School of Electrical Engineering, KAIST 
    }
\begin{document}
%
\maketitle
%
\input{tex/00_Abstract.tex}

\input{tex/01_Introduction.tex}

\input{tex/02_0_Related_Works}

\input{tex/03_0_Methodology}

\input{tex/04_0_Experiments_and_Results}

\input{tex/05_Conclusion.tex}

\section{Acknowledgements}
This work was supported by Institute of Information \& communications Technology Planning \& Evaluation\,(IITP) grant funded by the Korea government\,(MSIT) [No.20220-00641, 50\%] and the National Research Foundation of Korea\,(NRF) grant funded by MSIT
[No. 2021R1A2C1014044, 50\%].

\vfill
\clearpage
\newpage
\pagebreak

\bibliographystyle{IEEEbib}
\bibliography{strings,refs}

\newpage
\appendix
\input{tex/10_Appendix}

\end{document}

%% file: command/mycommand.tex
\usepackage{microtype}
\usepackage[T1]{fontenc}
\usepackage{wrapfig,lipsum,booktabs}

\usepackage{soul}
\usepackage{dsfont}
\usepackage{enumerate}
\usepackage{enumitem}
\usepackage{kotex}
\usepackage{hyperref}
\usepackage{amsmath}
\usepackage{amsfonts}
\usepackage{bbm}
\usepackage{dsfont}
\usepackage[Symbol]{upgreek}
\usepackage{lscape}
\usepackage{caption}
\usepackage{balance}
\usepackage{xspace}
\usepackage{float}
\usepackage{soul}
\usepackage{wasysym}
\usepackage{multirow}
\usepackage{array, boldline, rotating}
\usepackage{makecell}
\usepackage{blindtext}

\usepackage{amssymb}
\usepackage{pifont}
\renewcommand*\eqref[1]{(\ref{#1})}

\newcommand{\eg}{\emph{e.g.,~}}
\newcommand{\ie}{\emph{i.e.,~}}

\definecolor{Tan}{RGB}{210,180,140}
\definecolor{LightCyan}{rgb}{0.88,1,1}
\definecolor{LightGray}{gray}{0.9}
\definecolor{goldenrod}{rgb}{1.0,0.84,0.3}
\definecolor{shadecolor}{named}{LightGray}
\usepackage{tablefootnote}

%% file: tex/00_Abstract.tex
\begin{abstract} 
Audio-visual speech recognition (AVSR) aims to transcribe human speech using both audio and video modalities. In practical environments with noise-corrupted audio, the role of video information becomes crucial. However, prior works have primarily focused on enhancing audio features in AVSR, overlooking the importance of video features. In this study, we strengthen the video features by learning three temporal dynamics in video data: context order, playback direction, and the speed of video frames. Cross-modal attention modules are introduced to enrich video features with audio information so that speech variability can be taken into account when training on the video temporal dynamics. Based on our approach, we achieve the state-of-the-art performance on the LRS2 and LRS3 AVSR benchmarks for the noise-dominant settings. Our approach excels in scenarios especially for babble and speech noise, indicating the ability to distinguish the speech signal that should be recognized from lip movements in the video modality. We support the validity of our methodology by offering the ablation experiments for the temporal dynamics losses and the cross-modal attention architecture design.
\end{abstract}

\begin{keywords}
robust audio-visual speech recognition, video temporal dynamics, cross-modal attention
\end{keywords}

%% file: tex/01_Introduction.tex
\section{Introduction}
\label{sec:intro}

Audio-visual speech recognition (AVSR)~\cite{noda2015audio, 7178347, afouras2018deep, ma2021end, shi2022learning, ma2023auto}
represents a paradigm, where the integration of both auditory and visual modalities plays a crucial role for advancing speech recognition capabilities.
This multimodal approach utilizes not only the auditory cues existing in speech but also valuable visual information, such as lip movements.
However, the conventional AVSR methods do not fully exploit the potential of visual information~\cite{xu2020discriminative, hong2022visual}, which becomes significant when the audio-only speech recognition system is susceptible to background noise\,\cite{ren2021learning, chen2023leveraging}.
In such practical scenarios, it is essential to allow the AVSR system to rely on video information rather than overly-corrupted audio information.

Previous studies\,\cite{hong2022visual, hu2023hearing, hu2023cross} have mainly focused on enhancing noisy audio features or reducing modality gap, whereas few works have explored directly enhancing the video features with video-oriented learning for AVSR.
In particular, the audio enhancement is performed by taking advantage of the undistorted video information \cite{xu2020discriminative, hong2022visual} or restoring clean audio by a viseme-to-phoneme cluster mapping \cite{hu2023hearing}.
Also, several studies have explored to fuse the audio and video features with cross-modal attention \cite{wei2020attentive, wang2024mlca} or have proposed a contrastive loss to minimize the discrepancy between the two modalities \cite{hu2023cross, zhang2022learning}.
While these methods can be considered to improve the performance of AVSR, they have not investigated the intrinsic characteristics of video modality, such as temporal dynamics \cite{liu2017robust, liu2018learning, yun2022time} and spatio-temporal correlation \cite{caballero2017real, yi2019progressive}.

\begin{figure}[!t]
    \centering
    \includegraphics[width=0.95\linewidth]{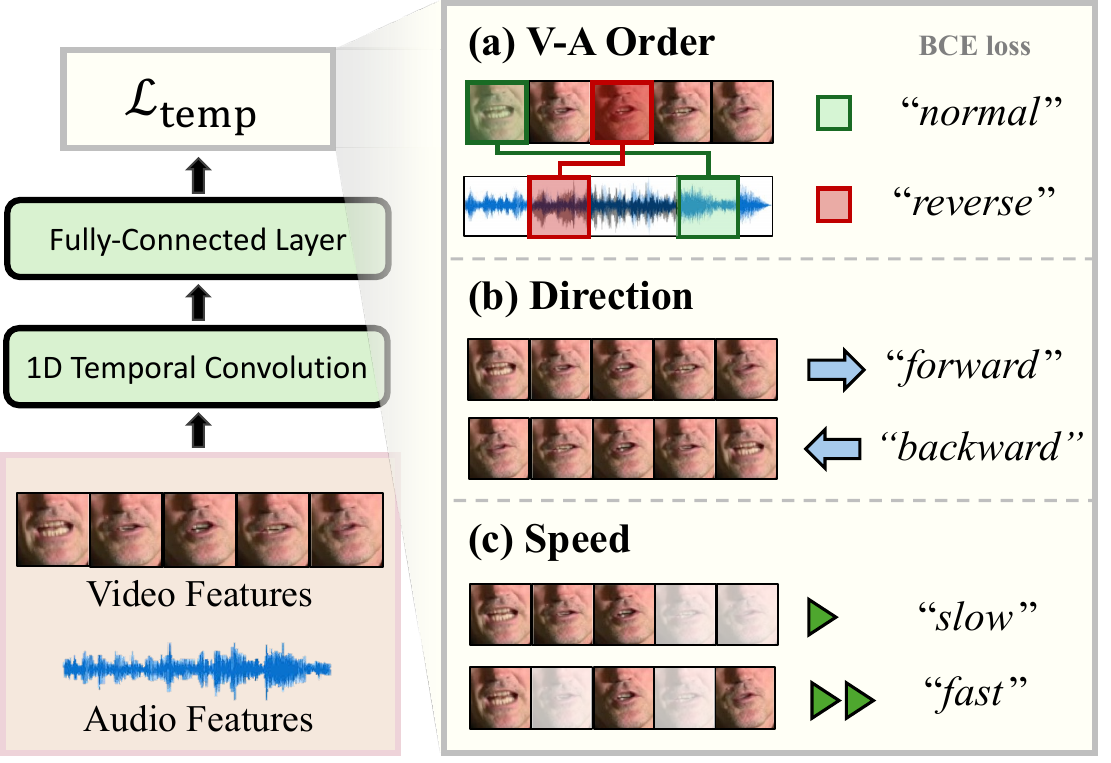}
    \vspace{-5pt}
    \caption{Our proposed temporal dynamics guidance\,($\mathcal{L}_\text{temp}$) involves predicting \textbf{(a)} the context order considering both video\,(\textbf{V}) and audio\,(\textbf{A}) modalities ($\mathcal{L}_\text{order}$; Eq.\,\ref{eq:order-loss}), \textbf{(b)} playback direction ($\mathcal{L}_\text{direction}$; Eq.\,\ref{eq:direction-loss}), and \textbf{(c)} whether certain frames are skipped or not ($\mathcal{L}_\text{speed}$; Eq.\,\ref{eq:speed-loss}). Each video temporal predictor is consisted of 1D convolution and fully-connected\,(FC) layers.}
    \label{fig:temp-dynamic-loss}
    \vspace{-5pt}
\end{figure}

In this work, we suggest training on temporal dynamics in the video data to enhance video features, making the AVSR system refer more to visual information.
Figure\,\ref{fig:temp-dynamic-loss} describes our training method in detail, where video features are processed through the temporal predictor to address each visual-related task.
Our method focuses on predicting three key objectives: (1) the context order between two random video and audio frames, (2) the playback direction of video frames, and (3) the playback speed of video frames.
As similar approaches have previously demonstrated success in the action recognition tasks \cite{yun2022time, jenni2023audio, dave2023no}, our AVSR system can discriminate the target speech to be recognized with the lip movement pair in the presence of multiple speakers.
Consequently, our video features are expected to encapsulate richer temporal understanding of the lip movements and audio context alignment, making the AVSR system more robust in the noisy audio condition.

To boost the effectiveness of learning temporal dynamics, we incorporate a cross-modal attention module within the video streamline, injecting audio information into the video features.
This structure enables video features to consider temporal variability in speech, such as coarticulation and variations in speaking speed, which can only be captured by attending multiple adjacent audio frames.
Understanding the temporal order of context between speech and lip movements also necessitates the cross-attention between two modalities.
Furthermore, to prevent misguidance of the temporal dynamics caused by distorted audio features, we implement an additional cross-modal attention into the audio streamline.
This attention module is trained with a refinement loss, utilizing clean video data to refine the noisy audio inputs.

To sum up, we propose a cross-modal attention structure to both video and audio modality streamlines, enhancing video and audio features with video temporal dynamics and audio refinement learning, respectively.
Our main contributions in this paper include the followings:
\vspace{-4pt}
\begin{itemize}[leftmargin=*]
\setlength\itemsep{-3pt}
    \item \textbf{Video temporal dynamics learning.}\quad We particularly enhance the video features for AVSR with the explicit goal of learning temporal dynamics, thereby significantly improving robustness in noisy audio conditions. To this end, we design cross-modal attention modules for enhancing the correlation between video and audio features.
    \item \textbf{Robust AVSR performance.}\quad Evaluating on the LRS2\,\cite{son2017lip} and LRS3\,\cite{afouras2018lrs3} AVSR benchmarks with the MUSAN noise\,\cite{snyder2015musan} added, our method achieves the state-of-the-art N-WER\footnote{N-WER denotes word error rate\,(WER) averaged across all 4 noise types and 5 signal-to-noise ratio\,(SNR) levels (refer to Section\,\ref{sec:implementation}).}\,\cite{shi2022robust} on both benchmarks. In particular for the LRS3 benchmark, our method outperforms UniVPM\,\cite{hu2023hearing} (5.2\%) with N-WER of 4.6\%.
    \item \textbf{Validation through ablation studies.}\quad
    We also investigate the validity of our methodology by offering the ablation experiments for the temporal dynamics losses and the cross-modal attention architecture design.
\end{itemize}

%% file: tex/02_0_Related_Works.tex
\section{Related Works}

\input{tex/02_1_AVSR.tex}
\input{tex/02_2_TDL.tex}

%% file: tex/02_1_AVSR.tex
\subsection{Audio-Visual Speech Recognition}
Recent AVSR works have focused on creating better audio-visual multi-modal representations via sophisticated training schemes or scaling up to larger datasets.
AV-HuBERT\,\cite{shi2022learning} learns to predict the cluster assignments of audio features for the masked prediction training.
Modality dropout is introduced for audio-visual fusion to prevent models from excessively relying on the audio modality.
Several approaches\,\cite{lian2023av, haliassos2022jointly, haliassos2024braven} employ a teacher-student framework, where the teacher model weights are updated via an exponential moving average of the student model weights, to predict contextualized target representations for the masked frames.
Auto-AVSR\,\cite{ma2023auto} incorporates the pretrained automatic speech recognition\,(ASR) model for creating pseudo-labels of the unlabeled video dataset.
The most recent finding\,\cite{chang2024conformer} illustrates that the linear projection is sufficient as a visual front-end with large-scale datasets.

Since speech recognition is often susceptible to background noise or ambient speech, addressing noise robustness in the AVSR system is a practical and important problem.
To this end, the follow-up study of AV-HuBERT\,\cite{shi2022robust} suggests to leverage noise-augmented audio for pretraining AV-HuBERT\,\cite{shi2022learning}.
UniVPM\,\cite{hu2023hearing} proposes a viseme-phoneme mapping to restore clean audio from lip movements under noisy environments.
Reinforcement learning is also utilized for robust AVSR by encouraging an agent to explore optimal strategies for WER\,\cite{chen2023leveraging}.
GILA\,\cite{hu2023cross} fuses audio and video representations with consecutive cross-modal attention blocks and implements contrastive loss to model the temporal consistency between audio-visual frames.
Our work departs from prior works in the aspect of enhancing video features through temporal dynamics learning. 
We propose integrating a cross-modal attention module into the existing AVSR system to enhance its robustness against various types of noise.

%% file: tex/02_2_TDL.tex
\subsection{Temporal Dynamics Learning}

Temporal dynamics refers to the information about changing patterns over multiple consecutive temporal frames, which has proven to help understanding videos, not necessarily with sound, in action recognition tasks\,\cite{yun2022time, dave2023no}.
The correlation between adjacent video frames can be boosted by learning temporal self-supervision tasks, such as predicting the direction of token's temporal flow\,\cite{yun2022time} or whether certain frames are skipped\,\cite{dave2023no}.
Recent work has extended temporal dynamics into a multi-modal scope, proposing an inter-modal contrastive loss that learns longer-term dynamics through context ordering between video and audio data\,\cite{jenni2023audio}.
Our work aims to enhance video features for noise-robust AVSR by training temporal dynamics with simple binary classification tasks in a self-supervised manner.
This approach differs from contrastive learning\,\cite{jenni2023audio}, which involves a complicated process of sampling positive and negative pairs and challenging optimization.

%

%% file: tex/03_0_Methodology.tex
\section{Methodology}
\label{sec:method}

\input{tex/03_1_Overview}

\input{tex/03_2_A2V}
\input{tex/03_3_V2A}

%% file: tex/03_1_Overview.tex
\subsection{Overview}

Our approach (Figure \ref{fig:temp-dynamic-loss} and \ref{fig:AXA-architecture}) is designed to reinforce the features of each modality, achieved by the temporal dynamics loss\,($\mathcal{L}_\text{temp}$) and the refinement loss\,($\mathcal{L}_\text{ref}$).
We insert a cross-modal attention structure between the front-end feature extractors and the pretrained AVSR encoder and train this attention structure with the two aforementioned losses.
In the video streamline, audio information is injected by the audio-to-video\,(A2V) cross-modal attention as a key and a value to train the temporal dynamics of video features.
Vice-versa, the video-to-audio\,(V2A) cross-modal attention utilizes clean video information as a key and a value to refine the noisy audio features and reduce the impact of noise on the subsequent A2V module.
We block the gradient flow from one side to the other for a stable training, ensuring that each loss can only train its corresponding streamline's cross-modal attention.
Consequently, the reinforced video and audio features are input to the pretrained AVSR encoder.

%% file: tex/03_2_A2V.tex
\subsection{A2V Video Temporal Dynamics Guidance}
\label{sec:A2V}
\noindent\textbf{A2V cross-modal attention.}\quad We propose enhancing video features by introducing a cross-modal attention structure, in which audio features serve as the key and the value, and video features serve as the query.
Rather than employing the simple fusion methods that enforce the alignment within a single frame, such as channel-wise concatenation or frame-wise addition\,\cite{ma2021end, burchi2023audio, petridis2018end}, we employ the attention mechanism to ensure that the video features take into account various speech context present in multiple audio frames. 
Furthermore, this A2V cross-modal attention module can help the video features comprehend temporal variability in speech, such as coarticulation or speaking speed, which can only be captured by attending multiple adjacent audio frames.

Let us denote the outputs of the video and audio front-ends as $\mathbf{f}_\text{v},\, \mathbf{f}_\text{a} \in \mathbb{R}^{T \times D}$, respectively.
They share the same sequence length $T$ and the channel dimension $D$. 
To inject audio information into the video features, we implement a stacked attention\footnote{Multi-head attention\,\cite{vaswani2017attention} mechanism is employed for the SA and CA module, but we omit its notation for brevity.} module; self-attention\,(SA) and then cross-modal attention\,(CA).
The SA module in the beginning is for preparing the features with referencing the other modality information.
$\mathbf{f}_\text{v}$ is first transformed by the query, key, and value matrices, $\mathbf{W}_q, \mathbf{W}_k, \mathbf{W}_v \in \mathbb{R}^{D \times D}$, followed by a SA mechanism.
Given a single FC layer $\mathbf{W}_{\text{fc}} \in \mathbb{R}^{D \times D}$,
\begin{align}
\label{eq:self-attention-video}
    \text{SA}(\mathbf{f}_\text{v}) &= \text{Attention}(\mathbf{f}_\text{v};\, \mathbf{f}_\text{v};\, \mathbf{f}_\text{v}) \nonumber \\
    &= \text{softmax}\left(\frac{(\mathbf{f}_\text{v}\mathbf{W}_{q})(\mathbf{f}_\text{v}\mathbf{W}_{k})^{\top}}{\sqrt{D}}\right)(\mathbf{f}_\text{v}\mathbf{W}_{v}), \\
    \mathbf{f}'_\text{v} &= \text{SA}(\mathbf{f}_\text{v}) \cdot \mathbf{W}_{\text{fc}_1}.
\end{align}

The resulting video features are then processed by the A2V cross-modal attention module, employing $\mathbf{f}'_\text{v} \in \mathbb{R}^{T \times D}$ as the query and the audio features $\tilde{\mathbf{f}}_\text{a}$ as the key and value.
Here, $\tilde{\mathbf{f}}_\text{a}$ is refined by another cross-modal attention in order to facilitate the accurate learning of temporal dynamics for the video features\,(refer to Section\,\ref{sec:V2A} for details on the audio feature refinement process).

\begin{align}
    \text{CA}(\mathbf{f}'_\text{v};\, \tilde{\mathbf{f}}_\text{a}) &= \text{Attention}(\mathbf{f}'_\text{v};\, \tilde{\mathbf{f}}_\text{a};\, \tilde{\mathbf{f}}_\text{a}), \\
    \tilde{\mathbf{f}}_\text{v} = \mathbf{f}_\text{v} &+ \text{CA}(\mathbf{f}'_\text{v};\, \tilde{\mathbf{f}}_\text{a}) \cdot \mathbf{W}_{\text{fc}_2}.
    \label{eq:cross-attention-video}
\end{align}
Our final audio-incorporated video features $\tilde{\mathbf{f}}_\text{v} \in \mathbb{R}^{T \times D}$ are residual summation of the original features and the FC layer's output of the A2V cross-modal attention.

\begin{figure}[!t]
    \centering
    \includegraphics[width=0.9\linewidth]{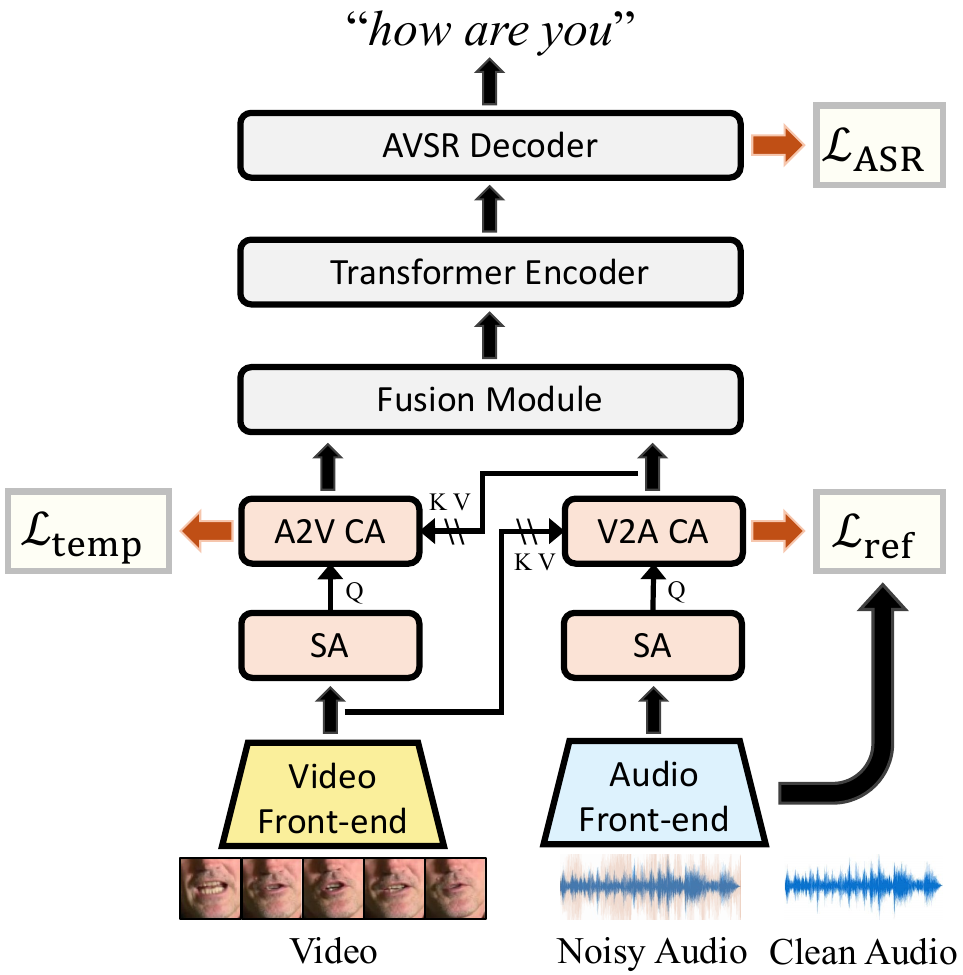}
    \vspace{-5pt}
    \caption{Our cross-modal attention structure is inserted between the feature extractors and the AVSR encoder. This structure leverages clean video to refine audio, and then learns video temporal dynamics given the refined audio features. Note that the gradient is not backpropagated between the two modalities so that training $\mathcal{L}_\text{temp}$ and $\mathcal{L}_\text{ref}$ does not interfere each other.}
    \label{fig:AXA-architecture}
    \vspace{-20pt}
\end{figure}

\vspace*{5pt}
\noindent\textbf{Temporal dynamics guidance.}\quad 
We train temporal dynamics on the video features, allowing them to be stronger contributors for AVSR.
The reliance on video information is more pronounced in AVSR with noisy audio conditions\,\cite{ ren2021learning, chen2023leveraging}, therefore, strengthening the video features would be essential.
Previous studies\,\cite{yun2022time, jenni2023audio, dave2023no} have shown improved performance in the action recognition tasks by learning video temporal dynamics. 
Viewing the AVSR task as continuous action recognition of lip movements, likewise, allows us to leverage temporal dynamics learning to understand natural lip movements.
For instance, discerning whether the given lip movement frames are being played forward or backward can help the video features be enriched with natural lip movements over time.

To synchronize the temporal positions of the context in video and audio, we involve the temporal order loss in a cross-modal way.
The order loss learns to predict the context order of two randomly selected frames, one from the video sequence and another from the audio sequence (refer to Figure \ref{fig:temp-dynamic-loss}(a)).
Let us denote $\tilde{\mathbf{f}}_\text{v} = (\tilde{v}_1, \cdots, \tilde{v}_T)^{\top}$ and $\tilde{\mathbf{f}}_\text{a} = (\tilde{a}_1, \cdots, \tilde{a}_T)^{\top}$ for the enhanced video and audio features, which are the outputs from each cross-modal attention module.
The context order loss $\mathcal{L}_\text{order}$ is defined as
\begin{equation}
\label{eq:order-loss}
    \mathcal{L}_\text{order} = \sum_{i,j, i \neq j} \text{BCE}(g(\tilde{v}_i \lVert \tilde{a}_j), y), 
\end{equation}
where $\lVert$ is a channel-wise concatenation.
The frame order labels are $y=1$ for $i<j$ and $y=0$ for $i>j$.
$\text{BCE}$ refers to a binary cross-entropy loss, and $g(\cdot)$\footnote{For the intuitive loss formulations, the inputs of the binary classifiers in Eqs.\,\ref{eq:order-loss}--\ref{eq:speed-loss}\,(\ie $\tilde{v}$ and $\tilde{a}$), are the features that have already undergone 1D temporal convolution in the binary classifier.} is a binary predictor network, composed of 1D temporal convolution\,\cite{bai2018empirical} followed by an FC layer.
The purpose of temporal convolution is to fuse temporally adjacent features, avoiding ambiguity where a single characteristic\,(\eg phoneme) may appear in multiple places within a sequence.

Furthermore, we propose the direction loss and speed loss to train temporal dynamics that are revealed in a short time duration, particularly learning the local temporal dynamics.
The direction loss predicts whether consecutive temporal length $t$ video frames are playing forward\,($y=1$) or backward\,($y=0$) (refer to Figure \ref{fig:temp-dynamic-loss}(b)).
\begin{align}
    \label{eq:direction-loss}
      \mathcal{L}_\text{direction} = \sum_{i}\,\, &\mathrm{BCE}(g(\tilde{v}_i \lVert \cdots \lVert \tilde{v}_{i+t-1}), 1) \nonumber \\
      &+ \mathrm{BCE}(g(\tilde{v}_{i+t-1} \lVert \cdots \lVert \tilde{v}_{i}), 0).
\end{align}
Additionally, the speed loss predicts whether a given video sequence is playing in a regular speed\,($y=1$) or skipping the frames in the speed of $k>1$\,($y=0$) (refer to Figure \ref{fig:temp-dynamic-loss}(c)).
\begin{align}
    \label{eq:speed-loss}
      \mathcal{L}_\text{speed} = \sum_{i}\,\, &\mathrm{BCE}(g(\tilde{v}_i \lVert \tilde{v}_{i+1} \lVert \cdots \lVert \tilde{v}_{i+t-1}), 1) \nonumber \\
      &+ \mathrm{BCE}(g(\tilde{v}_i \lVert \tilde{v}_{i+k} \lVert \cdots \lVert \tilde{v}_{i+(t-1)k}), 0).
\end{align}
These two loss functions help the video features learn how lip shape moves naturally over a short duration of time.
We also mark that the predictors are not shared across the different temporal dynamics losses and discarded for inference.

%% file: tex/03_3_V2A.tex
\subsection{V2A Audio Refinement}
\label{sec:V2A}

We have suggested the audio information be injected into the video features by the A2V cross-modal attention for training the video temporal dynamics.
However, perturbed audio information with noise may lead misguidance in learning the temporal dynamics.
Similar to previous AVSR works\,\cite{xu2020discriminative, hong2022visual, hu2023hearing} that have performed audio enhancement with accompanying clean video information, we aim to refine the audio inputs through a V2A cross-modal attention.

Analogous to the video streamline\,(Eq.\,\ref{eq:self-attention-video}--\ref{eq:cross-attention-video}), audio features are modified by the stacked SA-CA module, where the V2A cross-modal attention is performed to refine the noisy audio features with the help of video features.
\begin{align}
    \label{eq:v2a-cross-attention}
    \mathbf{f}'_\text{a} &= \text{SA}(\mathbf{f}_\text{a}) \cdot \mathbf{W}_{\text{fc}_3}, \\
    \tilde{\mathbf{f}}_\text{a} &= \mathbf{f}_\text{a} + \text{CA}(\mathbf{f}'_\text{a}; \mathbf{f}_\text{v})  \cdot \mathbf{W}_{\text{fc}_4}. \label{eq:asym-a2v}
\end{align}
We remind this audio refinement process in Eq.\,\ref{eq:asym-a2v} is preceding the video feature enhancement in Eq.\,\ref{eq:cross-attention-video}.
The modified audio features $\tilde{\mathbf{f}}_\text{a}$ reference the clean audio features $\mathbf{f}_\text{a, clean}$, with loss calculated by the mean-squared error\,(MSE).
The clean audio features are not processed further through our audio streamline.
\begin{equation}
    \label{eq:audio-denoising-loss}
    \mathcal{L}_\text{ref} =  \lVert\tilde{\mathbf{f}}_\text{a} - \mathbf{f}_\text{a, clean}\rVert^2.   
\end{equation}

\vspace{5pt}
\noindent\textbf{Overall training loss.}\quad
The reinforced audio and video features are input to the AVSR encoder, optimizing the entire model with the sequence-to-sequence ASR loss\,\cite{shi2022learning}, $\mathcal{L}_\text{ASR}$. 
Our final loss is the linear combination of each loss as follows:
\begin{align}
    &\mathcal{L}_\text{temp} = \mathcal{L}_\text{order} + \mathcal{L}_\text{direction} + \mathcal{L}_\text{speed}, \\
    &\mathcal{L} = \mathcal{L}_\text{ASR} + \lambda_\text{temp} \,\mathcal{L}_\text{temp} + \lambda_\text{ref} \,\mathcal{L}_\text{ref},
\end{align}
where $\lambda_\text{temp}$ and $\lambda_\text{ref}$ are the coefficients for video temporal loss and audio refinement loss.
We highlight that our method improves the performance of AVSR by simply adding losses during the fine-tuning stage, thus, does not require substantial training cost for pretraining.

%% file: tex/04_0_Experiments_and_Results.tex
\section{Experiments and Results}
\label{sec:exp}

\input{tex/04_1_Implementation_Details}
\input{tex/04_2_Main_Results}

\input{tex/04_3_Ablations}

%% file: tex/04_1_Implementation_Details.tex
\subsection{Implementation Details}
\label{sec:implementation}

\input{tex_table/main_result_lrs3.tex}

\textbf{Datasets.}\quad
We perform our experiments on the AVSR benchmarks, LRS2\,\cite{son2017lip} and LRS3\,\cite{afouras2018lrs3}.
Most of our experimental configurations follow prior works\,\cite{hu2023hearing, shi2022robust}, which include the extraction and addition of noise from MUSAN\,\cite{snyder2015musan}\,(\textit{babble}, \textit{music}, and \textit{natural}) and LRS3\,(\textit{speech}) datasets.
For training, we sample 0\,dB SNR noise and always add it to the clean speech signal for training.
We use noise from the MUSAN test set for evaluation as done in \cite{chen2023leveraging, hu2023hearing, hu2023mir}, as well as the babble noise synthesized from LRS3\,\cite{shi2022learning, xu2020discriminative}.
We report WER evaluated on noise-perturbed test set with 5 different SNR levels: $\{-10, -5, 0, 5, 10\}$.
The evaluation metric is N-WER\,\cite{shi2022robust}\,(AVG), the average WER across 4 noise categories and 5 SNRs.
We also report noise-dominant N-WER\,(N\,$\ge$\,S), which only considers 3 non-positive SNR levels.

\vspace*{5pt}
\noindent\textbf{Model and training description.}\quad
We adopt AV-HuBERT-\textsc{Large} model \cite{shi2022robust} as our backbone.
As an initialization, we load the pretrained checkpoint from\,\cite{shi2022robust}, pretrained on noise-augmented LRS3\,\cite{afouras2018lrs3}\,+\,VoxCeleb2\,\cite{chung2018voxceleb2}, and then fine-tune the model for 60K steps on the LRS2 or LRS3 train set.
For the first 48K steps, we freeze the AVSR encoder and front-ends of both modalities while training the AVSR decoder, the stacked SA-CA modules, and temporal predictors.
We adopt negative log-likelihood for $\mathcal{L}_\text{ASR}$.
$\lambda_\text{temp}$ and $\lambda_\text{ref}$ are 0.05 and 0.1, respectively.
We use a kernel size of 3 for the 1D convolution in the binary temporal classifiers and set the temporal length $t$ as 3 in Eqs.\,\ref{eq:direction-loss} and \ref{eq:speed-loss}.

%% file: tex_table/main_result_lrs3.tex
\begin{table*}[!t]
    \centering
    \small
    \caption{Comparisons of WER\,(\%) with our model and prior works on the LRS3\,\cite{afouras2018lrs3} AVSR benchmark. PT Type denotes whether the AVSR encoder is pretrained with noise-augmented audio. For evaluation, noise is sampled from the MUSAN\,\cite{snyder2015musan} dataset, while the results with babble noise from LRS3 are marked \textcolor{olive}{green}. We average the WER for music and natural noises\,\cite{hu2023hearing, shi2022robust}. We cite AV-HuBERT\,\cite{shi2022robust} results from the appendix of the original paper, which match the N-WER results\,(6.9\% and 5.8\%).
    }
    \label{tab:main_result_lrs3}
    \vspace{-5pt}
    \addtolength{\tabcolsep}{-2pt}
    \renewcommand{\arraystretch}{1.1}
    \resizebox{\textwidth}{!}{
    \begin{tabular}{lc|cccccc|cccccc|cccccc|cc|c}
    \toprule
    \multirow{2}{*}{Method} & \multirow{2}{*}{\!PT Type\!} & \multicolumn{6}{c|}{Babble, SNR (dB) $=$} & \multicolumn{6}{c|}{Speech, SNR (dB) $=$} & \multicolumn{6}{c|}{Music\,+\,Natural, SNR (dB) $=$} & \multicolumn{2}{c|}{N-WER} & Clean \\
    & & -10 & -5 & 0 & 5 & 10 & \!\textbf{AVG}\! & -10 & -5 & 0 & 5 & 10 & \textbf{AVG} & -10 & -5 & 0 & 5 & 10 & \textbf{AVG} & AVG & N\,$\ge$\,S & $\infty$ \\
    \midrule
    TM-seq2seq~\cite{afouras2018deep} & - & - & - & \textcolor{olive}{42.5} & - & - & - & - & - & - & - & - & - & - & - & - & - & - & - & - & - & 7.2 \\
    EG-seq2seq~\cite{xu2020discriminative} & - & \textcolor{olive}{38.6} & \textcolor{olive}{31.1} & \textcolor{olive}{25.5} & \!\textcolor{olive}{24.3}\! & \!\textcolor{olive}{20.7}\! & \!\textcolor{olive}{28.0}\! & - & - & - & - & - & - & - & - & - & - & - & - & - & - & 6.8 \\
    GILA-Conformer~\cite{hu2023cross}\!\! & - & - & - & - & - & - & - & - & - & - & - & - & - & - & - & - & - & - & - & 7.0 & - & 2.0 \\
    \midrule
    AV-HuBERT~\cite{shi2022robust} & clean & \textcolor{olive}{30.0} & \textcolor{olive}{15.2} & \textcolor{olive}{5.9} & \textcolor{olive}{2.7} & \textcolor{olive}{1.9} & \textcolor{olive}{11.1} & 15.9 & 7.5 & 3.9 & 2.4 & 1.9 & 6.3 & 12.1 & 5.9 & 3.1 & 2.2 & 1.8 & 5.0 & \textcolor{olive}{6.9} & \textcolor{olive}{10.0} & 1.4 \\
    UniVPM~\cite{hu2023hearing} & clean & 28.1 & 13.8 & 5.1 & 2.2 & 1.7 & 10.2 & 14.5 & 6.7 & 3.3 & 2.1 & 1.7 & 5.7 & 10.7 & 5.2 & 2.7 & 1.9 & 1.6 & 4.4 & 6.2 & 9.1 & 1.2 \\
    \cellcolor[HTML]{f0f3f4}\textbf{Ours} & \cellcolor[HTML]{f0f3f4}{clean} & \cellcolor[HTML]{f0f3f4}\thead{\small \textcolor{olive}{28.3}\\\small 24.8} & \cellcolor[HTML]{f0f3f4}\thead{\small \textcolor{olive}{13.4}\\\small 11.2} & \cellcolor[HTML]{f0f3f4}\thead{\small \textcolor{olive}{\,\,4.8\,\,} \\\small \,\,4.6\,\,} & \cellcolor[HTML]{f0f3f4}\thead{\small \textcolor{olive}{\,\,2.4\,\,} \\\small \,\,2.3\,\,} & \cellcolor[HTML]{f0f3f4}\thead{\small \textcolor{olive}{\,\,1.7\,\,} \\\small \,\,1.9\,\,} & \cellcolor[HTML]{f0f3f4}\thead{\small \textbf{\textcolor{olive}{\,\,10.1\,\,}} \\\small \textbf{\,\,9.0\,\,}} & \cellcolor[HTML]{f0f3f4}9.9 & \cellcolor[HTML]{f0f3f4}5.2 & \cellcolor[HTML]{f0f3f4}3.4 & \cellcolor[HTML]{f0f3f4}2.3 & \cellcolor[HTML]{f0f3f4}1.6 & \cellcolor[HTML]{f0f3f4}\textbf{4.5} & \cellcolor[HTML]{f0f3f4}9.7 & \cellcolor[HTML]{f0f3f4}4.9 & \cellcolor[HTML]{f0f3f4}2.6 & \cellcolor[HTML]{f0f3f4}2.0 & \cellcolor[HTML]{f0f3f4}1.8 & \cellcolor[HTML]{f0f3f4}\textbf{4.2} & \cellcolor[HTML]{f0f3f4}\thead{\small \textbf{\textcolor{olive}{\,\,\,5.7\,\,\,}}\\\small \textbf{\,\,\,5.4\,\,\,}} & \cellcolor[HTML]{f0f3f4}\thead{\small \textbf{\textcolor{olive}{\,\,\,8.3\,\,\,\,}}\\\small \cellcolor[HTML]{f0f3f4}\textbf{\,\,\,7.8\,\,\,\,}} & \cellcolor[HTML]{f0f3f4}1.5 \\
    \midrule
    u-HuBERT~\cite{hsu2022u} & noisy & - & - & {4.1} & - & - & - & - & - & - & - & - & - & - & - & - & - & - & - & - & - & 1.2 \\
    AV-HuBERT~\cite{shi2022robust} & noisy & \textcolor{olive}{28.4} & \textcolor{olive}{13.4} & \textcolor{olive}{5.0} & \textcolor{olive}{2.6} & \textcolor{olive}{1.9} & \textcolor{olive}{10.3} & 11.4 & 4.6 & 2.9 & 2.2 & 1.8 & 4.6 & 9.7 & 4.7 & 2.5 & 1.9 & 1.8 & 4.1 & \textcolor{olive}{5.8} & \textcolor{olive}{8.3} & 1.4 \\
    MIR-GAN~\cite{hu2023mir} & noisy & - & - & - & - & - & - & - & - & - & - & - & - & - & - & - & - & - & - & 5.6 & - & 1.2 \\
    UniVPM~\cite{hu2023hearing} & noisy & 26.8 & 12.1 & 4.0 & 2.1 & 1.6 & 9.3 & 10.4 & 4.1 & 2.5 & 2.0 & 1.6 & 4.1 & 8.7 & 4.1 & 2.1 & 1.7 & 1.5 & \textbf{3.6} & 5.2 & 7.5 & 1.2 \\
    MSRL~\cite{chen2023leveraging} & noisy & 22.4 & 11.3 & 4.5 & 2.3 & {-} & {-} & 7.2 & 3.4 & 2.3 & 1.8 & - & - & 8.5 & 4.3 & 2.4 & 1.7 & - & - & - & 6.8 & 1.3 \\
    \rowcolor[HTML]{f0f3f4}
    \cellcolor[HTML]{f0f3f4}\textbf{Ours} & \cellcolor[HTML]{f0f3f4}noisy & \cellcolor[HTML]{f0f3f4}\thead{\small\textcolor{olive}{25.8}\\\small22.7} & \cellcolor[HTML]{f0f3f4}\thead{\small\textcolor{olive}{11.9}\\\small9.9} & \thead{\small\textcolor{olive}{\,\,4.4\,\,}\\\small \,\,4.0\,\,} & \thead{\small\textcolor{olive}{\,\,2.4\,\,}\\\small \,\,2.2\,\,} & \cellcolor[HTML]{f0f3f4}\thead{\small \textcolor{olive}{\,\,1.8\,\,}\\\small\,1.8\,}&\cellcolor[HTML]{f0f3f4}\thead{\small\textbf{\textcolor{olive}{\,\,\,9.3\,\,\,}}\\\small\textbf{\,\,\,8.1\,\,\,}} & \cellcolor[HTML]{f0f3f4}5.4 & 3.2 & 2.5 & 1.8 & 1.8 & \textbf{2.9} & 8.7 & 3.7 & 2.4 & 2.0 & 1.7 & 3.7 & \thead{\small \textcolor{olive}{\textbf{\,\,\,4.9\,\,\,}}\\\small \textbf{\,\,\,4.6\,\,\,}} & \thead{\small \textcolor{olive}{\textbf{\,\,\,6.9\,\,\,\,}}\\\small \textbf{\,\,\,6.4\,\,\,\,}} & 1.5 \\
    \bottomrule
    \end{tabular}
    }
\end{table*}

%% file: tex/04_2_Main_Results.tex
\subsection{Robust AVSR Benchmark Results}
\label{sec:results}

In Table\,\ref{tab:main_result_lrs3}, we present the AVSR performance of our proposed method, evaluated on the LRS3\,\cite{afouras2018lrs3} benchmark.
Our model consistently surpasses AV-HuBERT\,\cite{shi2022robust} across all four noise types, as indicated by N-WER of 5.7\% and 4.9\%, depending on whether the encoder is pretrained with noise-augmented audio\,(\ie PT Type).
Also, it outperforms MIR-GAN\,\cite{hu2023mir} and UniVPM\,\cite{hu2023hearing} by 5.4\%/4.6\% N-WER for clean/noisy PT Type, respectively, attaining a new state-of-the-art performance.
Our methods especially excelling in babble and speech noise while offering comparable results in music and natural noises.
This highlights the importance of learning video temporal dynamics with audio information, rendering the AVSR model to accurately distinguish the target speech signal in multi-speaker scenarios by attending lip movements in the video data.
Meanwhile, it is crucial to acknowledge a trade-off in noise robustness.
Our method, catered to noise-corrupted conditions, leads to exceptional performance gain in such scenarios but a slight degradation in the clean speech setting, which has been similarly observed in other noise-robust ASR works\,\cite{wang2022wav2vec, wang2023hubert, ng2023hubert}.

For the noise-dominant N-WER\,(N\,$\ge$\,S), our results exhibit great effectiveness in certain scenarios, highlighting its robustness and real-world applicability.
The comparisons with recent works, including MSRL\,\cite{chen2023leveraging} and UniVPM\,\cite{hu2023hearing}, substantiate that our method bolsters the robustness of our AVSR system with achieving 7.8\% and 6.4\% noise-dominant N-WER regard to PT Type.
Our method also outperforms AV-HuBERT\,\cite{shi2022robust} for both PT Type with 8.3\% and 6.9\%.
In contrast to previous approaches that rely on the general methods for reducing the modality gap, such as contrastive learning\,\cite{hu2023cross} or adversarial learning\,\cite{hu2023hearing, hu2023mir}, our method incorporates the inherent characteristics of video features, thereby resulting in superior performance for the noise-dominant setting.
Importantly, refined audio information is injected into video features at this stage, making our cross-modal attention design well-suited for the noise-robust AVSR task.

The trend of the aforementioned results continues in the LRS2\,\cite{son2017lip} benchmark\,(Table\,\ref{tab:main_result_lrs2}).
Our method surpasses the recent noise-robust AVSR works, MIR-GAN\,\cite{hu2023mir} and UniVPM\,\cite{hu2023hearing}, with 5.9\% N-WER and 7.7\% noise-dominant N-WER.
Comparing with AV-HuBERT\,\cite{shi2022robust}, around 11\% of relative performance gain is achieved in both average and noise-dominant N-WER.
This confirms that our method is still effective, regardless of the difference between the pretraining and fine-tuning datasets.

\input{tex_table/main_result_lrs2.tex}

%% file: tex_table/main_result_lrs2.tex
\begin{table*}[!t]
    \centering
    \small
    \caption{Comparisons of WER\,(\%) with our model and prior works on the LRS2\,\cite{son2017lip} AVSR benchmark. For the AV-HuBERT\,\cite{shi2022robust} results, we fine-tune the pretrained AV-HuBERT encoder on LRS2.
    All models are pretrained with noise-augmented audio\,(\ie PT Type is noisy) except for GILA-Conformer\,\cite{hu2023cross}. For evaluation, augmented noise is sampled from the MUSAN\,\cite{snyder2015musan} dataset, while the results with babble noise from LRS3 are marked \textcolor{olive}{green}. We average the WER for music and natural noises\,\cite{hu2023hearing, shi2022robust}.
    }
    \label{tab:main_result_lrs2}
    \vspace{-5pt}
    \addtolength{\tabcolsep}{-0.5pt}
    \renewcommand{\arraystretch}{1.1}
    \resizebox{\textwidth}{!}{
    \begin{tabular}{l|cccccc|cccccc|cccccc|cc|c}
    \toprule
    \multirow{2}{*}{Method} & \multicolumn{6}{c|}{Babble, SNR (dB) $=$} & \multicolumn{6}{c|}{Speech, SNR (dB) $=$} & \multicolumn{6}{c|}{Music\,+\,Natural, SNR (dB) $=$} & \multicolumn{2}{c|}{N-WER} & Clean \\
    & -10 & -5 & 0 & 5 & 10 & \textbf{AVG} & -10 & -5 & 0 & 5 & 10 & \textbf{AVG} & -10 & -5 & 0 & 5 & 10 & \textbf{AVG} & AVG\! & N\,$\ge$\,S & $\infty$ \\
    \midrule
    AV-HuBERT~\cite{shi2022robust} & \textcolor{olive}{31.7} & \textcolor{olive}{15.1} & \textcolor{olive}{6.3} & \textcolor{olive}{4.1} & \textcolor{olive}{3.2} & \textcolor{olive}{12.1} & 8.6 & 5.5 & 4.2 & 3.7 & 3.3 & 5.1 & 11.0 & 6.0 & 4.3 & 3.4 & 3.0 & 5.5 & \textcolor{olive}{7.1} & \textcolor{olive}{9.5} & 2.6 \\
    GILA-Conformer~\cite{hu2023cross} & - & - & - & - & - & - & - & - & - & - & - & - & - & - & - & - & - & - & 11.2 & - & 3.1 \\    
    MIR-GAN~\cite{hu2023mir} & - & - & - & - & - & - & - & - & - & - & - & - & - & - & - & - & - & - & 7.0 & - & 2.2 \\
    UniVPM~\cite{hu2023hearing} & {30.1} & {13.7} & {5.7} & {4.1} & {3.2} & {11.4} & 7.5 & 5.1 & 3.4 & 3.1 & 2.8 & \textbf{4.4} & 10.9 & 5.0 & 3.8 & 3.1 & 2.8 & \textbf{5.1} & 6.5 & 8.7 & 2.2 \\
    \rowcolor[HTML]{f0f3f4}
    \textbf{Ours} & \thead{\small \textcolor{olive}{27.8}\\\small{22.4}} & \thead{\small \textcolor{olive}{12.6}\\\small{10.1}} & \thead{\small \textcolor{olive}{5.2}\\\small{5.0}} & \thead{\small \textcolor{olive}{3.7}\\\small{3.7}} & \thead{\small \textcolor{olive}{3.0}\\\small{3.2}} & \thead{\small \textbf{\,\,\textcolor{olive}{10.4}\,\,}\\\small{\,\,\textbf{8.9}}\,\,} & 7.5 & 4.7 & 3.8 & 3.1 & 2.9 & \textbf{4.4} & 9.9 & 6.0 & 3.8 & 3.3 & 2.9 & 5.2 &  \thead{\small \textbf{\,\,\textcolor{olive}{6.3}\,\,}\\\small \,\,{\textbf{5.9}}\,\,} & \thead{\small \textbf{\,\,\,\textcolor{olive}{8.4}\,\,\,}\\\small {\,\,\,\textbf{7.7}\,\,\,}} & 2.7 \\
    \bottomrule
    \end{tabular}
    }
\end{table*}

%% file: tex/04_3_Ablations.tex
\subsection{Ablation Study}
\label{sec:ablation-loss}

\textbf{Training losses.}\quad
Table\,\ref{tab:ablation_loss} shows our investigation on the independent effects of the video temporal learning and audio refinement losses, by systematically adding each loss component.
The plain ASR loss\,($\mathcal{L}_\text{ASR}$) matches the ASR fine-tuning loss of AV-HuBERT\,\cite{shi2022learning} baseline.
Starting from this, we find that utilizing the video temporal losses\,($\mathcal{L}_\text{ASR}$ + $\mathcal{L}_\text{temp}$) plays a crucial role in improving overall performance, verifying the importance of strengthening the video features for robust AVSR.
Employing the audio refinement loss\,($\mathcal{L}_\text{ASR}$ + $\mathcal{L}_\text{ref}$) also shows the performance improvement but not as much as video temporal losses.
Combining all the proposed losses\,($\mathcal{L}_\text{ASR}$ + $\mathcal{L}_\text{temp}$ + $\mathcal{L}_\text{ref}$), we achieve the best AVSR performance which shows refining noise-perturbed audio is particularly crucial for correctly guiding the temporal dynamics to video features.

We further examine the performance of visual speech recognition\,(VSR) to demonstrate the enhancement of visual features without any audio information.
For the VSR evaluation, the audio features are input as zero vectors, and all the other implementations remain the same.
Our method, which is accompanied with the video temporal learning, produces a lower 32.5\% WER compared to the baseline\,(33.7\%), indicating its better representations of lip movements.
However, when video temporal learning is missing or has been misguided by noisy audio, the VSR performance is adversely affected, resulting in 33.2\% and 33.3\% WER, respectively.

\vspace*{5pt}
\noindent\textbf{Temporal dynamics loss.}\quad
In Table\,\ref{tab:ablation_loss_architecture}(a), we further investigate how each type of temporal dynamics loss affects the AVSR results. 
Since our video temporal loss consists of three losses with various combinations possible, we exclude each one individually to understand its impact on the total temporal loss.
We observe a performance drop when one of these functions is omitted from the total temporal dynamics loss, especially noticeable when the context order loss or speed loss is not included.
Additionally, we include a video-to-video order loss, which predicts the order of randomly selected two video frames.
This strategy has not gained improvement, suggesting that learning the video-to-audio order loss implicitly encompasses learning the video order itself.

\vspace*{5pt}
\noindent\textbf{Attention architecture designs.}\quad
Table\,\ref{tab:ablation_loss_architecture}(b) demonstrates the effectiveness of our cross-modal architecture design.
As described in Figure\,\ref{fig:AXA-architecture}, we use a stacked SA-CA layer for each video and audio streamline.
The ablation experiments illustrate the necessity of both SA and CA layers, with the CA layer revealed to be the most crucial component.
This underscores the significance of attending the other modality for learning video temporal dynamics or refining noisy audio.
We also replace CA with the second SA layer to compare models with same number of parameters, which is proved to be sub-optimal.

%

%% file: tex/05_Conclusion.tex
\section{Conclusion}
\label{sec:conc}

In this paper, we have proposed to train the temporal dynamics of video features and employ the cross-modal attention, for the noise-robust AVSR system.
Our temporal dynamics learning includes predicting the context order between two random audio and video frames in each sequence, the playback direction, and the playback speed of video frames.
Of our stacked SA-CA module, the cross-modal attention plays a crucial role for correlating each modality to one another.
Our methodology achieves the state-of-the-art performance on the LRS2 and LRS3 AVSR benchmarks, particularly when the input audio is perturbed with various noise types and SNR levels.
This illustrates the significance of enhancing the video features based on the temporal dynamics of lip movements.
By extensive ablation studies about our suggested loss and attention architecture, we have confirmed the video temporal dynamics learning with cross-modal attention design is essential for improving the noise-robustness of AVSR system.

\input{tex_table/ablation_loss}

\input{tex_table/ablation_tdl_arch}

%% file: tex_table/ablation_loss.tex
\begin{table}[!t]
    \centering
    \small
    \caption{Ablation experiments for the proposed training loss functions.
    For AVSR, we average the WER results\,(\%) across noise-dominant settings\,(N\,$\ge$\,S) on the LRS3\,(L) and MUSAN\,(M) babble noise.
    VSR is evaluated with video-only inputs, discarding the audio modality.}
    \vspace{-5pt}
    \addtolength{\tabcolsep}{-1pt}
    \label{tab:ablation_loss}
    \resizebox{0.8\linewidth}{!}{
    \begin{tabular}{l|c|c|c}
    \toprule
    Loss & AVSR\,(L) & AVSR\,(M) & VSR \\
    \midrule
    $\mathcal{L}_\text{ASR}$ & 15.6 & 14.6 & 33.7 \\
    $\mathcal{L}_\text{ASR}$ + $\mathcal{L}_\text{ref}$ & 14.8 & 12.9 & 33.2 \\
    $\mathcal{L}_\text{ASR}$ + $\mathcal{L}_\text{temp}$ & 14.3 & 12.5 & 33.3 \\
    $\mathcal{L}_\text{ASR}$ + $\mathcal{L}_\text{temp}$ + $\mathcal{L}_\text{ref}$ (\textbf{ours}) & \textbf{14.0} & \textbf{12.2} & \textbf{32.5} \\
    \bottomrule
    \end{tabular}
    }
    \vspace{2pt}
\end{table}

%% file: tex_table/ablation_tdl_arch.tex
\begin{table}[!t]
    \centering
    \small
    \caption{Ablation experiments for (a) the proposed temporal dynamics loss functions and (b) the attention module architecture designs.
    For evaluation, we average the WER results across three SNRs (N\,$\ge$\,S) on the MUSAN babble noise.
    }
    \vspace{-5pt}
    \addtolength{\tabcolsep}{2pt}
    \label{tab:ablation_loss_architecture}
    \renewcommand{\arraystretch}{1.1}
    \resizebox{0.518\linewidth}{!}{
    \begin{tabular}{l|c}
    \toprule
    Loss & \!\!\!\!WER\,(\%)\!\!\!\! \\
    \midrule
    $\mathcal{L}_\text{ASR}$ + $\mathcal{L}_\text{temp}$ + $\mathcal{L}_\text{ref}$ (\textbf{ours})\!\! & \textbf{12.2} \\
    ~~~(-) video-to-audio order & 12.7 \\
    ~~~(-) direction & 12.5 \\
    ~~~(-) speed & 12.9 \\
    ~~~(+) video-to-video order & 12.4 \\
    \bottomrule
    \multicolumn{2}{c}{(a) Temporal dynamics loss ablation} \\
    \end{tabular}
    }
    \hfill
    %
    \resizebox{0.465\linewidth}{!}{
    \begin{tabular}{l|c}
    \toprule
    \multicolumn{2}{l}{\textcolor{gray}{(Loss fixed as: $\mathcal{L}_\text{ASR}$\,+\,$\mathcal{L}_\text{temp}$\,+\,$\mathcal{L}_\text{ref}$)}}\!\! \\
    \midrule
    Architecture & WER\,(\%) \\
    \midrule
    SA\,+\,CA (\textbf{ours}) & \textbf{12.2} \\
    ~~~(--) SA & 12.4 \\
    ~~~(--) CA & 13.0 \\
    SA\,+\,SA & 12.5 \\
    \bottomrule
    \multicolumn{2}{c}{(b) Attention architecture ablation} \\
    \end{tabular}
    }
\end{table}

%% file: tex/10_Appendix.tex
\onecolumn
\begin{center}
    \textbf{\Large Appendix}
\end{center}

\section{Implementation Details}

\subsection{Datasets}
\label{sec:appx_dataset}

We perform our experiments on LRS2\,\cite{son2017lip} and LRS3\,\cite{afouras2018lrs3}, datasets comprising around 224 and 433 hours of audio-visual speech data, respectively, from over 5,000 speakers.
Most of our experimental configurations follow\,\cite{hu2023hearing} and\,\cite{shi2022robust}, including the noise augmentation and evaluation protocol.
We extract noise from MUSAN\,\cite{snyder2015musan}\,(\textit{babble}, \textit{music}, and \textit{natural}) and LRS3\,(\textit{speech}) datasets, and partition them into train, validation, and test sets.
For training, we sample noise at 0\,dB SNR and always add it to the clean speech signal.
For evaluation, we use noise from the MUSAN test set, as done in \cite{chen2023leveraging, hu2023hearing, hu2023mir}, as well as synthesizing the babble noise by randomly mixing 30 audio clips from LRS3, following \cite{shi2022learning, xu2020discriminative}.

\subsection{Model and Training Description}

We adopt AV-HuBERT-\textsc{Large} model \cite{shi2022robust} as our backbone, which consists of 24 and 9 Transformer\,\cite{vaswani2017attention} blocks as the AVSR encoder and decoder, respectively.
This backbone consists of a feature extractor for each modality, where the visual feature extractor is a modified ResNet-18, and the audio feature extractor is a linear projection layer. Extracted features are concatenated to form fusion audio-visual features, which are input to the Transformer encoder.
While more recent AVSR models exist\,\cite{ma2023auto, lian2023av, haliassos2024braven}, we apply our method to AV-HuBERT for fair comparison with previous works\,\cite{chen2023leveraging, hu2023hearing, shi2022robust, hu2023mir} that utilize the same noise-augmenting protocol and pretrained AVSR encoder.
The hidden dimension of the cross-modal attention is 1024, same as that of the AVSR encoder.

As an initialization, we load the pretrained checkpoint from\,\cite{shi2022robust}, pretrained on noise-augmented LRS3\,\cite{afouras2018lrs3}\,+\,VoxCeleb2\,\cite{chung2018voxceleb2}, and then fine-tune the model for 60K steps on the LRS2 or LRS3 train set.
We observe unstable fine-tuning due to the insertion of untrained cross-modal attention module between pretrained AVSR encoder and front-ends of both modalities.
Therefore, we freeze all pretrained modules for the first 48k steps which means only cross-modal attention module and the AVSR decoder are trained during these steps.
We adopt negative log-likelihood loss for $\mathcal{L}_\text{ASR}$ in a sequence-to-sequence manner with a Transformer decoder.
The total number of parameters in our whole model is 500M while AV-HuBERT-\textsc{Large} is 477M, implying the stacked SA-CA module only accounts for less than 5\%.
Our code is implemented upon the \texttt{fairseq} pipeline.

For the binary temporal classifier, we use a kernel size of 3 for the 1D convolutional layer followed by the single FC layer.
We set the temporal length $t$ as 3, sampling 3 consecutive frames to formulate direction and speed loss functions.
A single frame in our model spans 40 ms, which is shorter than the average English phoneme duration of 100 ms.
Thus, we set the kernel size as 3 to match between lip movements and acoustic characteristics such as phonemes, while also minimizing ambiguity from repeated characteristics in a single sequence. Our internal experiments with kernel sizes of 5 and 7 have yielded similar performance to kernel size 3, while leading to slightly increased computational overhead. However, one can consider using a larger kernel size when working with a language that has longer duration phonemes than English.

\subsection{Positional Encoding}

The conventional Transformer architecture consists of positional encoding (PE) layers prior to self-attention layers to provide contextual information regarding the position of tokens within a sequence. In our framework, which is focused on learning temporal dynamics, we aim to prevent learning redundant information attributable to the PE layer. Consequently, in our cross-modal modules, we apply PE to only the keys and values derived from the other modality streamline, excluding the queries in the CA layer. This implementation ensures that the PE layer does not directly learn temporal dynamics on the video streamline side, where the queries are primarily tasked with computing the temporal losses. For the same reason, SA layers in our cross-modal module are designed without incorporating any PE layers.